\documentclass[final,5p,times,compress,twocolumn]{elsarticle}

\usepackage{amssymb}
\usepackage{amsmath}
\usepackage{booktabs}
\usepackage{xcolor}
\usepackage{array}
\usepackage{bigints}
\usepackage{orcidlink}
\usepackage{hyperref}
\usepackage{bm}
\usepackage[normalem]{ulem}
\usepackage{enumitem}

\journal{Physics Letters B}

\begin{document}

\begin{frontmatter}



\title{Bayesian evaluation of hadron-quark phase transition models through neutron star observables in light of nuclear and astrophysics data}

\author[BITS_Hyd]{Debanjan Guha Roy} 
\author[BITS_Hyd]{Anagh Venneti} 
\author[CFisUC]{Tuhin Malik} 
\author[BITS_Hyd]{Swastik Bhattacharya}\ead{swastik@hyderabad.bits-pilani.ac.in} 
\author[BITS_Hyd]{Sarmistha Banik} 
\affiliation[BITS_Hyd]{organization={Department of Physics, BITS Pilani, Hyderabad Campus}, 
            city={Hyderabad},
            postcode={500078}, 
            state={Telangana},
            country={India}}
\affiliation[CFisUC]{organization={CFisUC, Department of Physics, University of Coimbra},
    city={Coimbra},
    postcode={3004-516},
    country={Portugal}}
            
\begin{abstract}
We investigate the role of hybrid and nucleonic equations of state (EOSs) within neutron star (NS) interiors using Bayesian inference to evaluate their alignment with recent observational data from NICER and LIGO-Virgo (LV) collaborations. We find that smooth hybrid EOSs are slightly favoured in explaining NS mass-radius relations, particularly for pulsars such as PSR J0030+0451 and PSR J0740+6620. However, this preference is not definitive, as gravitational wave (GW) data does not significantly differentiate between our hybrid and nucleonic models. Our analysis also reveals tensions between older NICER data and recent measurements for PSR J0437$-$4715, highlighting the need for more flexible EOS models. Through two sampling approaches --- one fixing the hadronic EOS set and the other without fixing the same, we demonstrate that the hybrid EOS model can incorporate stiffer EOSs, resulting in a better agreement with NICER data but leading to higher tidal deformability, which is less consistent with GW observations.
In some recent publications a parameter $d_c$, related to the trace anomaly and its derivative, is used to indicate the presence of deconfined quark matter.
We find that our hadronic model, which does not include phase transition to deconfined matter, under the influence of imposed constraints, is able to predict values below 0.2 for $d_c$ at around five times saturation density. The hybrid model goes below this threshold at lower densities under the same conditions. 
\end{abstract}


\end{frontmatter}



\section{Introduction}
Recent observational data, particularly from gravitational wave (GW) detections and precise pulsar measurements, have significantly advanced our understanding of neutron star (NS) interiors.
Mass-radius estimation by NICER (Neutron Star Interior Composition Explorer) reported a radius of \(12.71_{-1.19}^{+1.14}\) km and a mass of \(1.34_{-0.16}^{+0.15}\) M$_\odot$ \cite{Riley_2019} whereas an alternative study\cite{Miller_2019} found a radius of \(13.02_{-1.06}^{+1.24}\) km and a mass of \(1.44_{-0.14}^{+0.15}\) M$_\odot$ for the pulsar PSR J0030+0451. Furthermore, NICER's observations for the pulsar PSR J0740+6620 indicated an equatorial circumferential radius of \(12.39^{+1.30}_{-0.98}\) km and a mass of \(2.072_{-0.066}^{+0.067}\) M$_\odot$ (with a 68\% confidence interval)\cite{Riley_2021}.  Most recent NICER observations of PSR J0437$-$4715  have measured its mass and radius 1.418 M$_\odot$ and $11.36_{-0.63}^{+0.95}$ km respectively\cite{Choudhury_2024}. 

In addition to astrophysical observations, numerous research efforts\cite{Tews_2013, Drischler_2020, Huth_2021, Malik_2023, Venneti_2024} have focused on determining the EOS of strongly interacting NS matter, considering the saturation properties of nuclear matter. Inferred mass-radius values from the detection of GWs from binary NS merger event GW170817\cite{LIGOScientific:2018cki, Bauswein:2017vtn, Annala:2017llu}, combined with ab initio calculations like perturbative quantum chromodynamics (pQCD) at very high densities\cite{Komoltsev2021jzg, Komoltsev_2024} and chiral effective field theory ($\chi$EFT) models of neutron matter at low densities, have provided additional insights into the EOS of dense matter. 

These combined empirical observations and theoretical frameworks are essential for advancing our understanding of nuclear matter under extreme conditions\cite{Nandi_2019, Biswas_2021, Ghosh_2022, Huth_2022, Imam:2024gfh}. Owing to the incredibly high density within the core of an NS, the highly compressed matter might contain strange particles like hyperons\cite{Banik_2014, Chatterjee_2016, Malik:2021nas}, deconfined quarks\cite{Weber_2005, Orsaria_2014}, Bose-Einstein condensates of antikaons\cite{Malik_2021}, etc.

In recent decades, numerous quark models (QM) have been investigated, suggesting the possibility of two-flavour colour superconductivity\cite{Alford:1997zt, Bonanno:2011ch, Ivanytskyi:2022oxv}, the colour-flavour locked (CFL) phase, and even more complex states such as the Larkin-Ovchinnikov-Fulde-Ferrell (LOFF)\cite{Mannarelli:2006fy, Rajagopal:2006ig}, and crystalline superconducting phases\cite{Haskell_2007}.
The MIT bag model\cite{Chodos:1974pn, Johnson:1978uy} and the Nambu-Jona-Lasinio (NJL) model are two widely used frameworks to describe QM within NS \cite{Schertler:1999xn, Steiner:2000bi, Buballa:2003qv, Contrera:2022tqh}.
Additionally, the hybrid EOS models often utilize the Mean Field Approximation of QCD (MFTQCD), which is derived from the QCD Lagrangian \cite{Fogaca:2010mf}. 
This approach considers the decomposition of a gluon field into soft and hard momentum components, where the mean field approximation for hard gluons contributes to stiffening the EOS, allowing for high pressures at high energy densities. On the other hand, the soft gluon fields generate condensates that soften the EOS, resulting in a bag-like term \cite{Fogaca:2010mf}. 
Usually, strange degrees of freedom soften the EOS, which results in a lower maximum mass and more compact star\cite{Weber_2005, Tolos:2020aln}. However, some model-independent research indicates that the cores of most massive stars may align with the presence of QM\cite{Alford_2019, Annala:2019puf, Annala:2023cwx}.

Besides mass-radius, $f-$mode oscillation frequency is a potential probe of stellar structure\cite{Andersson_1998, Kumar:2023rut}. $f-$mode oscillations are sensitive to EOS. Hence, they can also give clues about the presence of exotic matter in the core of NS\cite{Pradhan_2024, Thakur_2024_arXiv}. The study of correlations among various aspects of stellar structure, such as mass, radius, tidal deformability, and $f-$mode frequencies, helps to infer properties that are otherwise difficult to measure directly. For NSs, the correlation between observed data with theoretical model parameters allows stringent limits on the EOS of dense matter.
  
Some recent works\cite{Tan_2020, Shahrbaf_2020, Yamamoto_2023} have analysed  EOSs with hadron-quark phase transitions keeping in mind the latest astrophysical observations. The correlation between the radii of two NSs, like PSR J0030+0451, and PSR J0740+6620, with masses differing by $\approx$ 0.6$M_\odot$ as measured by NICER, was found to be noteworthy for EOSs with nucleonic degrees of freedom \cite{Essick_2023}. However, this correlation weakens during a first-order phase transition \cite{Lin_2023}. As a result, it might serve as a signature for the existence of hadron-quark phase transitions in NSs. In a recent publication \cite{GuhaRoy_2024}, the correlation in the radius domain to $f-$mode oscillation frequency domain is translated using semi-universal relations. Also, analysis based on a few selected hybrid EOSs from the CompOSE database\cite{Typel_2015} with NICER data hints at non-nucleonic degrees of freedom for matter inside the NS core \cite{GuhaRoy_2024}. 

In this letter, we perform a detailed statistical analysis using large sets of hybrid EOSs.
Considering the latest astrophysical constraints, we investigate whether NSs with a hadron-quark phase transition are more favourable compared to stars composed solely of nucleonic matter. 
Using recent observational data, we choose a systematic Bayesian inference approach to evaluate the Bayes Log Evidence for these two scenarios.
Specifically, we calculate the Bayes Log Evidence based on astrophysical data for three cases: a purely nucleonic EOS, a hybrid EOS utilizing NICER data for PSR J0030+0451 and PSR J0740+6620, and another hybrid EOS incorporating additional mass-radius constraints from the recent observation of PSR J0437$-$4715\cite{Choudhury_2024}. 

The article is organized as follows: Section \ref{sec:hyb_EOS_construction} discusses the construction of hybrid EOS. Section \ref{sec:oscillations} reviews the non-radial oscillations of NS.
In Section \ref{sec:results}, we present our results.
Finally, Section \ref{sec:conclusions} summarizes our findings and their implications for the study of NS  interiors.

\section{Construction of hybrid equations of state} 
\label{sec:hyb_EOS_construction}
We model both hadronic and quark components and employ a formalism to simulate smooth phase transitions. Our formalism includes sharp transitions as a special case.

\subsection{Hadronic and quark EOS}
We utilize the RMF model, including multiple non-linear meson interaction terms as done in Ref. \cite{GuhaRoy_2024} to construct the nucleonic segment of the EOS. Several constraints are imposed within the Bayesian framework --- low-density pure neutron matter constraints from $\chi$EFT calculations, constraints on a few nuclear matter saturation properties such as binding energy per nucleon, incompressibility for symmetric nuclear matter, nuclear symmetry energy, etc. 

We obtain the quark EOSs within the formalism of Mean Field Theory of Quantum Chromodynamics (MFTQCD) \cite{Albino:2024ymc}.
The MFTQCD EOS model is considered an improved version of the original MIT bag model. MIT Bag model with finite bag constant describes the dynamics of free massless quarks under the influence of a soft gluon background. However, the original version of the MIT bag model is incapable of modelling massive NSs \cite{Torres_2013}, which are of interest in astrophysics. MFTQCD can successfully reproduce quark stars with maximum masses beyond 2 $M_\odot$ and acceptable radii ($<12$ km) \cite{Franzon:2012in}.
We use the analytical expressions of Ref. \cite{Albino_2022} for pressure and energy density.
Please refer to \ref{sec:hadron_quark_EOS} for more details on the models used.

\subsection{Phase transition}
Some recent works like \cite{Kapusta:2021ney, Constantinou:2023ged} have constructed the hadron-quark phase transition as a smooth crossover transition by using a switching function that varies with one thermodynamic variable like the baryonic chemical potential ($\mu_B$) as one moves toward higher densities. The resulting transitions are between the two extremes: Maxwell and Gibbs construction. Maxwell construction imposes charge neutrality locally and leads to an abrupt jump in the energy density, whereas the Gibbs construction addresses charge neutrality on a global scale and results in smoother phase transitions\cite{glendenning}.

We use the expressions for pressure, energy density and baryon number density from Refs. \cite{Albino_2022, Hama_2006} to ensure a phase transition framework that includes the Maxwell construction as a special case depending on the values of the free parameters.
The two free parameters, $\delta_0$ and $\mu_C$ control the smoothness of the phase transition;
$\delta_0=0$ (MeV/fm$^3$)$^2$ and/or $\mu_C \approx 0 - 200$ MeV correspond to a sharp discontinuity in energy density; higher values of $\delta_0$ result in EOSs with a smoother transition \cite{Albino_2022}.
For the present study, we set the value of these parameters to {$\delta_0=50$, $\mu_C=700 \text{ MeV}$} to perform a quicker and less resource-intensive analysis (see \ref{sec:var_PT_params}).

We find the onset of phase transition from the intersection of the hadronic and the quark EOS in the $P-\mu_B$ plane. Below the threshold chemical potential $\mu_{B,tr}$ at the intersection, we have a pure hadronic phase.
We ensure the phase transition sets in well below the central density corresponding to the maximum mass NS for a given EOS and restrict the number density at the transition ($\rho_{B,tr}$) between 1.2 $\rho_{B,0}$ and 2.5 $\rho_{B,0}$ (see \ref{sec:PT_llhood}), where $\rho_{B,0} =$ 0.16 fm$^{-3}$ is the is the saturation number density.

\subsection{Bayesian approach for generating hadronic and hybrid EOS sets} 
\label{subsec:Bayes}

We have developed sets of NS matter EOSs to study NS properties within a Bayesian inference framework, incorporating the astrophysical constraints, namely the mass-radius measurement posterior from NICER for PSR J0030+0451, PSR J0740+6620, and PSR J0437$-$4715 along with the tidal deformability posterior of the two companions in the binary system involved in GW170817 (see the Table \ref{sec:loglikelihoods} for details).

The hybrid EOS is generated in two distinct ways: (i) \textbf{Fixed Hadronic (FH) scenario} --- by fixing the NL EOS distribution, i.e. sampling quark EOS parameters alongside a random NL EOS from the posterior of an EOS set featuring nucleonic degrees of freedom, and (ii) \textbf{Combined parameter sampling (CPS) scenario} --- by varying all parameters of both the NL model (RMF) and the quark model (MFTQCD) within the Bayesian framework. The second approach is computationally expensive due to the random combinations of model parameters from the NL and MFQCD, which could result in a null EOS if no transition occurs. 
Additionally, we apply pQCD-derived constraints on our hybrid EOSs at $\rho_B = 7 \times \rho_{B,0}$ as a likelihood function available at the URL\cite{Gorda_2023_link}. Please refer to the supplementary material for more details on nuclear data and the corresponding likelihood functions used in this work.

We configured the nested sampling algorithm ---\texttt{PyMultiNest} \cite{Buchner_2014} with 2000 live points. This setup enabled us to achieve a reliable posterior distribution with around 10,000 samples for each set.
In summary, we have created sets with around 10,000 EOSs for various scenarios as follows:
\begin{itemize}
   \item i) \emph{Set0}: exclusively hadronic.
   \item ii) \emph{Set1} FH: hybrid EOS set formed by fixing the hadronic profile from \emph{Set0}
   \item iii) \emph{Set2} FH: similar to \emph{Set1} FH, including constraints from PSR J0437$-$4715.
   \item iv) \emph{Set1} CPS: hybrid EOS  with combined parameter sampling from both NL and MFQCD.
   \item v) \emph{Set2} CPS: similar to \emph{Set1} CPS,  with additional constraints from PSR J0437$-$4715.
\end{itemize}
Each set aims to investigate different NL and QM combinations under various observational restrictions.

\section{Non-radial perturbations of spherically symmetric metric}
\label{sec:oscillations}
We study the structure of the spherically symmetric, static NSs with all the above EOS sets by solving the Tolman-Oppenheimer-Volkoff equations. We use the equilibrium mass $M$, corresponding radius $R$, variation of pressure, and mass inside the star to calculate the $f-$mode (\emph{fundamental} mode) frequencies with the orbital angular momentum number $l=2$ for a given mass of non-rotating NS.

We employ the formalism of non-radial stellar perturbations of non-rotating NS as given in \cite{Lindblom_1983}.
Assuming the perturbations to have a harmonic time dependence $\sim \exp(i \omega t)$, where $\omega$ is the oscillation frequency for a given mode, one can derive a set of four coupled differential equations for four perturbation functions with respect to the radial coordinate $r$\cite{GuhaRoy_2024}.
We solve these equations inside the star and match them at the star's surface with the perturbation functions obtained by solving the Zerilli equation\cite{Fackerell_1971} outside the star.

The solution of the system of differential equations with suitable boundary conditions is an eigenvalue problem in the frequency domain, with $\omega$ being the complex eigenvalue. The real part of $\omega$ gives us the oscillation frequency of the mode, while the inverse of the imaginary part represents the GW damping time of the mode.

\begin{figure*}[ht]
    \centering
    \includegraphics[width=\linewidth]{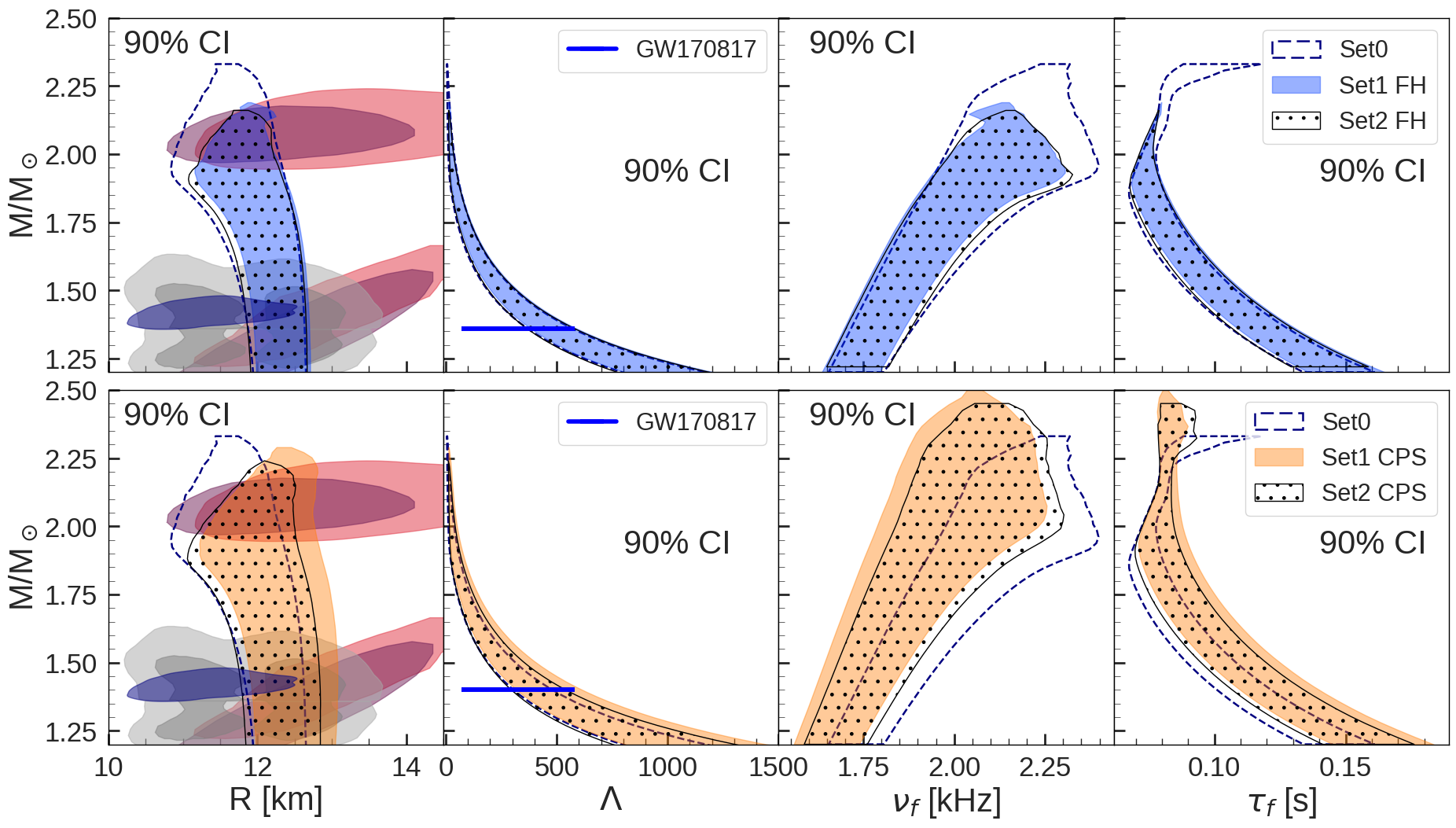}
    \caption{The 90\% credible interval (CI) regions for the NS mass-radius relationship $P(R|M)$, mass-tidal deformability $P(\Lambda|M)$, mass-$f-$mode frequency $P(\nu_f|M)$, and mass-damping time $P(\tau_f|M)$ for $f$ mode oscillations are depicted, based on the posterior distributions for \emph{Set0}, \emph{Set1} FH, and \emph{Set2} FH, in the top panels from left to right. The lower panels show the same distributions for \emph{Set1} CPS and \emph{Set2} CPS, along with \emph{Set0}. Different observational constraints for mass-radius and mass-tidal deformability are also plotted for comparison. In the mass-radius panel, the grey region signifies the constraints obtained from the binary components of GW170817, displaying their 90\% and 50\% credible intervals. Furthermore, the plot includes the 1$\sigma$ (68\%) CI for the 2D mass-radius posterior distributions of the millisecond pulsars PSR J0030+0451  \cite{Riley_2019, Miller:2019cac} and PSR J0740+6620 \cite{Riley:2021pdl, Miller:2021qha}, derived from NICER X-ray data. The most recent NICER measurements for the mass and radius of PSR J0437$-$4715 \cite{Choudhury_2024} are also presented. In the mass-tidal deformability panels, the horizontal bar represents the tidal deformability constraints for a 1.36 $M_\odot$ NS, obtained from the GW170817 merger event\cite{LIGOScientific:2018cki}.}
    \label{fig:MR_f_tau_set0}
\end{figure*}

\section{Results}
\label{sec:results}
In this section, we compare the results of the quantities like mass(M), radius(R), tidal deformability($\Lambda$), $f-$mode oscillation frequency($\nu_f$), and GW damping time($\tau_f$) calculated for the nucleonic and hybrid EOS sets within the two scenarios mentioned in Sec. \ref{subsec:Bayes}.
We check for the presence of conformal matter by evaluating a quantity $d_c$ as defined in \cite{Annala:2023cwx}, which depends on the trace anomaly and its derivative. Further, we compare the EOS models by determining the Bayes factor and also present the distribution of a quantity D$_L$ as done in \cite{Lin_2023}.

In Fig. \ref{fig:MR_f_tau_set0}, we show the 90\% credible intervals for the NS mass-radius $P(R|M)$, mass-tidal deformability $P(\Lambda|M)$, mass-$f-$mode frequency $P(\nu_f|M)$, and mass-damping time $P(\tau_f|M)$ of $f-$mode oscillations sequentially from left to right. Here, $P(A|B)$ stands for the conditional probability distribution of the quantity $A$ given $B$. The top panels depict intervals for the EOSs in the posterior distributions of \emph{Set0}, \emph{Set1} FH, and \emph{Set2} FH. The lower panels display the same distributions for \emph{Set0}, \emph{Set1} CPS and \emph{Set2} CPS. As listed in Sec. \ref{subsec:Bayes}, \emph{Set0} is obtained exclusively using nucleonic degrees of freedom, while the other sets include hybrid EOS, featuring a hadron-quark phase transition. The hybrid EOSs shown in the top panels are constructed by maintaining a constant set of nucleonic EOS --- the parameters of the MFTQCD model are sampled along with a random choice of nucleonic EOS from the posterior distribution of \emph{Set0}. The hybrid EOSs in the bottom panel are obtained by simultaneously sampling all the parameters of the RMF and MFTQCD models during the inference process. The distribution of NS properties for the hybrid EOSs in \emph{Set1} FH and \emph{Set2} FH is encompassed by that for the nucleonic \emph{Set0}, evident in the top panel of Fig. \ref{fig:MR_f_tau_set0}. Nevertheless, there is a significant decrease in the maximum mass and a marginal effect on the lower bounds of NS properties. \emph{Set0} serves as a baseline, with the median values --- $M_{\text{max}} = 2.07 \, M_\odot$, $R_{1.4} = 12.26$ km, and $\Lambda_{1.4} = 389$ (see Table \ref{tab:post_CPS}). We note that these values are within the expected ranges for typically observed values. \emph{Set1} FH exhibits slightly lower $M_{\text{max}}$ at 2.01 $M_\odot$, with marginally higher $R_{1.4}$ and $\Lambda_{1.4}$ values. This indicates EOSs which are stiffer in low densities and softer at high densities relative to the EOSs in \emph{Set0}. Imposition of additional constraints from PSR J0437$-$4715 in \emph{Set2} FH, leads to a reduction of approximately $\sim$ 0.1 km in $R_{1.4}$ and a slight decrease in $\Lambda_{1.4}$, and a minor decrease of 0.01 M$_\odot$ in the maximum NS mass compared to \emph{Set1} FH.

In the lower panels of Fig. \ref{fig:MR_f_tau_set0}, a slight reduction in the NS maximum mass is noticed for the hybrid sets --- \emph{Set1} CPS and \emph{Set2} CPS. However, there is a notable increase in the upper 90\% credible interval (CI) for other NS properties, including radius, tidal deformability, $f-$mode oscillation frequency, and GW damping time. \emph{Set1} CPS presents the highest $R_{1.4}$ (= 12.51 km) and $\Lambda_{1.4}$ (= 447), indicating a significantly stiffer EOS, leading to a larger and less compact NS. $M_{\text{max}}$ is slightly higher than that of \emph{Set0}, at 2.21 $M_\odot$. \emph{Set2} CPS has a $M_{\text{max}}$ similar to \emph{Set1} FH but shows slightly reduced $R_{1.4}$ and $\Lambda_{1.4}$ compared to \emph{Set1} CPS, indicating a less stiff EOS but still stiffer than those in \emph{Set0} and \emph{Set1} FH.
Overall, the differences across the sets reflect variations in the underlying EOSs: \emph{Set1} CPS and \emph{Set2} FH generally predict stiffer NS structures, while \emph{Set0} and \emph{Set1} FH suggest typical NS properties within the expected observational ranges.

In the two rightmost panels in Fig. \ref{fig:MR_f_tau_set0}, we find that $f-$mode oscillation frequencies $\nu_{f_{1.4}}$ reduces from 1.812 kHz for \emph{Set0} to 1.777 kHz for \emph{Set1} FH and 1.712 kHz for \emph{Set1} CPS. This echoes the trend in $R_{1.4}$ over the EOS sets under consideration --- it was observed in Ref. \cite{GuhaRoy_2024} that $\nu_{f_{1.4}}$ is inversely proportional to $R_{1.4}$. EOSs softer at high densities, therefore, will lead to larger NSs with lower values of $\nu_{f_{1.4}}$. Median values of 1.793 kHz and 1.744 kHz for $\nu_{f_{1.4}}$ corresponding to \emph{Set2} FH and \emph{Set2} CPS, respectively, then indicate softer EOSs. The median value of GW damping time $\tau_{f_{1.4}}$ is equal to 0.111 s \emph{Set0} and remains practically unchanged over the two hybrid EOS sets within the FH scenario. But it increases to 0.124 s and 0.120 s for \emph{Set1} and \emph{Set2} within the CPS scenario.

We assess the impact of NICER and GW data on the log-likelihood values of different EOS sets: \emph{Set0}, \emph{Set1} FH, \emph{Set2} FH, \emph{Set1} CPS, and \emph{Set2} CPS (see Fig. \ref{fig:log_likelihoods} in supplementary material). Our analysis reveals that the exclusively nucleonic EOS \emph{Set0}, consistently shows the highest log-likelihood peaks across all data scenarios (NICER, GW, and NICER+GW) compared to the other sets. In contrast, the hybrid EOS sets (\emph{Set1} FH and \emph{Set2} FH), where the nucleonic profile remains unchanged, demonstrated lower peaks. For the CPS scenario, which involves the simultaneous sampling of all parameters in the hybrid system, \emph{Set1} CPS exhibits a lower log-likelihood peak than \emph{Set0} for NICER data, but the opposite is true for GW data. However, when combining NICER and GW data, the total log-likelihoods are similar across these sets. This suggests that NICER data tend to favour the hybrid sets, whereas GW data lean towards the nucleonic set. The difference is attributed to the CPS scenario where the stiff nucleonic EOSs play a significant role; these EOSs were excluded during sampling in \emph{Set0} due to constraints from pQCD or NICER data for NSs with a mass of 2.08 $M_\odot$. After a phase transition, these EOSs become softer at high densities, although they are actually stiff at lower densities. Additionally, including PSR J0437$-$4715 data in \emph{Set2} FH or \emph{Set2} CPS results in higher log-likelihoods with NICER data, indicating potential tensions between the older NICER data and the new PSR J0437$-$4715 data, or a mismatch with the hybrid EOS model. This calls for the need to explore more adaptable models that can reconcile these differences.

For better clarity, we have computed the Bayes factor and the Log evidence for all the obtained posteriors for individual NICER, GW, and NICER+GW data (refer to the supplementary material). Comparing \emph{Set0} against \emph{Set1} FH and \emph{Set1} CPS, the evidence across different datasets (NICER, GW, and NICER+GW) consistently indicate no substantial preference, with Log Bayes factors suggesting very weak or nonexistent evidence favouring one over the other. We interpret the factors using Jeffrey's scale (see \ref{tab:Jeffreys} for the ranges).
Specifically, for NICER and NICER+GW, the Log Bayes factors are close to zero or slightly positive, implying minimal to no advantage of \emph{Set0} over \emph{Set1} FH and \emph{Set1} CPS. The evidence is weaker in the GW dataset, with no significant difference between \emph{Set0} and \emph{Set1} within either FH or CPS. In contrast, the results are more definitive when we compare \emph{Set2} FH and \emph{Set2} CPS against the other sets. For both NICER and NICER+GW, there is substantial evidence favouring \emph{Set0} over \emph{Set2} within both FH and CPS, as well as evidence favouring \emph{Set1} FH over \emph{Set2} CPS. Additionally, strong evidence supports \emph{Set1} CPS over \emph{Set2} FH. However, it remains very weak in the GW dataset, with no substantial differences detected between \emph{Set2} FH or CPS and the other sets.

\begin{figure}[h]
    \centering
    \includegraphics[width=\linewidth]{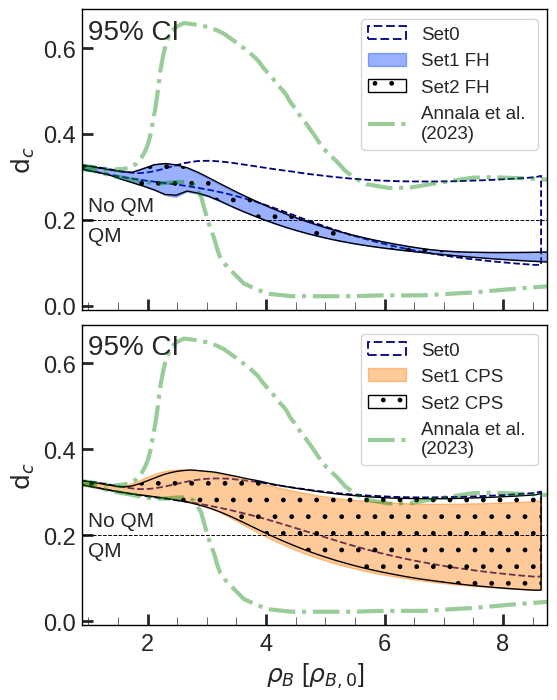}
    \caption{Plot of conformality factor $d_c$ as a function of baryon number density $\rho_B$. Annala et al. in Ref. \cite{Annala:2019puf} delineates the demarcation between conformal and non-conformal regions by $\text{d}_c = 0.2$. Notice our nucleonic EOSs cross the line at around 5 times the nuclear saturation density $\rho_{B,0}$.}
    \label{fig:dc_plot}
\end{figure}

The conformality parameter $d_c$ is given as $d_c = \sqrt{\Delta^2 + (\Delta')^2}$ in Ref. \cite{Annala:2023cwx}. Here, $\Delta \equiv \frac{1}{3} - \frac{p}{\epsilon}$ represents the renormalized trace anomaly mentioned in \cite{Fujimoto:2022ohj} and $\Delta' \equiv d \Delta / d \log \epsilon$, where $\epsilon$ is the energy density. $\Delta'$ is related to the squared speed of sound $c_s^2$ and $\gamma = \frac{d \ln p}{d \ln \epsilon}$ as $\Delta' = c_s^2 \left(\frac{1}{\gamma} - 1\right)$.
In the conformal limit, the values of $c_s^2$ and $\gamma$ approach $1/3$ and $1$, respectively. It is suggested in Ref. \cite{Annala:2023cwx} that $d_c \lesssim 0.2$ indicates a closeness to the conformal limit, as this would necessitate small values for both $\Delta$ and its derivative. Given that QM is expected to exhibit approximate conformal symmetry, a small $d_c$ could signal its presence.

In Figure \ref{fig:dc_plot}, we present the 90\% credible interval (CI) of the critical density, $d_c$, plotted against the baryon density, $\rho_B$. The top panel shows the results for \emph{Set0}, \emph{Set1} FH, and \emph{Set2} FH, while the bottom panel includes the same for \emph{Set0}, \emph{Set1} CPS, \emph{Set2} CPS. Here, the regions for the three sets are denoted by dashed, filled and dotted patches, respectively, in both the panels.
We compare these regions with those reported by Annala et al. in \cite{Annala:2023cwx} marked within the dash-dotted contour. \emph{Set0} exhibits a marginal bump much smaller compared to the significant one shown by Annala et al. at approximately 3 $\rho_{B,0}$. Moreover, in the nucleonic EOS posterior from \emph{Set0}, $d_c$ dips below the threshold of 0.2 at around 5 $\rho_{B,0}$. In contrast, the hybrid sets in the top panel display a narrower $d_c$ posterior that peaks slightly earlier, around 2.5 $\rho_{B,0}$, and drops below 0.2 noticeably sooner than in \emph{Set0}. This narrower distribution is attributed to the fixed nucleonic distribution within the hybrid EOS, allowing only the stiff components of the nucleonic EOS to contribute. Within the CPS scenario, where all parameters of the hybrid system are independently varied, the $d_c$ posterior is wider compared to that in the FH case. Here, the bump is slightly larger; it occurs at 2.5 $\rho_{B,0}$ earlier than that of \emph{Set0}, and falls below the critical value sooner. The entirety of the posterior is encompassed by the region obtained by Annala et al., indicating that the model used by Annala et al. is more flexible and allows for greater freedom than the approach we use in our study.\\

\begin{table}[h]
    \centering
    \begin{tabular}{ccccccc}
         \toprule
         Data & \multicolumn{2}{c}{$R_{1.34} - R_{2.07}$} & \multicolumn{2}{c}{$R_{1.418} - R_{2.07}$} & \multicolumn{2}{c}{$R_{1.44} - R_{2.07}$}\\
         \midrule
         \emph{Set0} & \multicolumn{2}{c}{0.77} & \multicolumn{2}{c}{0.80} & \multicolumn{2}{c}{0.81}\\
         \midrule
         & FH & CPS & FH & CPS & FH & CPS\\
         \midrule
         \emph{Set1} & 0.45 & 0.70 & 0.50 & 0.73 & 0.51 & 0.74\\
         \emph{Set2} & 0.56 & 0.66 & 0.61 & 0.70 & 0.62 & 0.71\\
         \bottomrule
    \end{tabular}
    \caption{Pearson correlation coefficient between radii of two NS with given masses for the sets of EOSs used in this work}
    \label{tab:rad_corr_new}
\end{table}

\begin{figure}
    \centering
    \includegraphics[width=\linewidth]{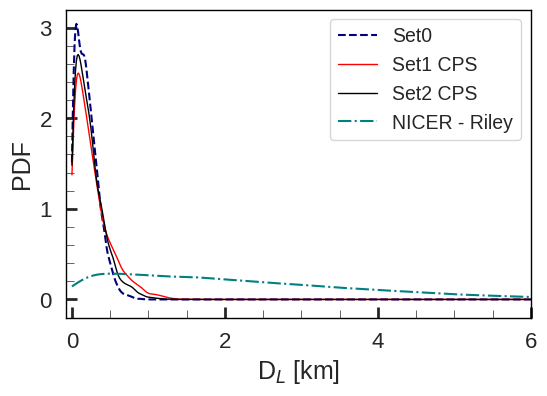}
    \caption{Probability density function (PDF) of D$_L$ values for nucleonic EOS set (\emph{Set0} --- navy blue dashed curve) and two hybrid EOS sets (\emph{Set1} --- red solid curve and \emph{Set2} --- black solid curve) within the combined parameter sampling (CPS) scenario. The teal dash-dotted curve represents the D$_L$ values for $R_{1.34}$ and $R_{2.07}$ from mass-radius posteriors for PSR J0030+0451\cite{Riley_2019} and  PSR J0740+6620\cite{Riley_2021}, respectively.}
    \label{fig:DL_plot}
\end{figure}

In Table \ref{tab:rad_corr_new}, we display the Pearson correlation coefficients between radii of NSs with masses: 1.34, 1.418, 1.44, and 2.07 M$_\odot$. These are the median values of mass for PSR J0030+0451\cite{Riley_2019}, PSR J0437$-$4715\cite{Choudhury_2024}, PSR J0030+0451\cite{Miller_2019}, and PSR J0740+6620\cite{Riley_2021} respectively. Notice how the correlation weakens more for the hybrid EOS sets within the FH scenario relative to the EOS sets within the CPS scenario.

In order to quantify the degree to which an EOS set is favoured by the NICER data based on the correlations between radii of two NSs with different masses, we compute the quantity D$_L$ following Ref. \cite{Lin_2023}. We determine a linear fit for the $R_{1.34}$ and $R_{2.07}$ with the nucleonic EOS set. D$_L$ is a measure of how much the point ($R_{1.34}$, $R_{2.07}$) for any given EOS deviates from the linear fit. With $R_{1.34}$ and $R_{2.07}$ values from the posterior (reported in \cite{Riley_2019, Riley_2021}) of NICER data for 1.34 M$_\odot$ pulsar and 2.07 M$_\odot$ pulsar, we calculate D$_L$ values for the NICER data.

In Fig. \ref{fig:DL_plot}, we compare probability density functions (PDF) of D$_L$ values for different EOS sets.
While prominent peaks occur at nearly the same D$_L$ value for our sets, the distribution for NICER data (teal-coloured dash-dotted curve) has a huge spread without any such clear peak. The spread of the dash-dotted curve points to the absence of a strong correlation between $R_{1.34}$ and $R_{2.07}$. This is expected since the calculation of D$_L$ values with NICER data involves data for two completely independent measurements of two different pulsars. At around $D_L = 0.6$ km, the PDF for our sets become comparable with the PDF for NICER data.
 
According to \cite{Lin_2023}, EOSs with sharp first-order phase transitions are needed to obtain such a weak correlation between the radii values. Evidently, the array of astrophysical data and nuclear physics constraints used in this work don't allow our hybrid EOS model to predict such EOSs.
Based on the evaluation of Kullback-Leibler (KL) divergence, the Kolmogorov-Smirnov (KS) test with the cumulative distribution functions (CDFs) (see \ref{sec:stat_tests}) and comparison of overlap area of the CDFs of the three abovementioned sets with respect to that of the NICER dataset, \emph{Set1} CPS emerges as the EOS set statistically closest to align with the NICER dataset used in this work.
\section{Conclusions}
\label{sec:conclusions}

In this study, we compare hybrid and nucleonic EOSs within NSs using Bayesian inference to evaluate their alignment with observational data.
Our focus is on hybrid EOS models incorporating hadron-quark phase transitions, irrespective of whether the transitions are sharp or smooth.
We assess their compatibility with recent mass-radius data from NICER and other astrophysical observables, including the tidal deformability measurements by the LIGO-Virgo collaboration.

To better understand the role of hybrid EOSs, we employ two sampling approaches: (i) fixing the distribution profile of the hadronic EOS and sampling random hadronic EOS from this set along with the quark EOS parameters (FH models) and (ii) implementing a comprehensive inference framework where all hadronic and quark EOS parameters are sampled together within the hybrid EOS system (CPS models). The latter approach, though computationally demanding, reveals that hybrid systems could incorporate stiff hadronic EOS models initially excluded from the purely nucleonic set.
These stiff EOSs undergo smooth phase transitions. They show a better fit with NICER data due to a shift in the mass-radius posterior with a marginally higher median value of radius at 1.4 $M_\odot$.
However, this shift results in higher tidal deformability values, which do not align well with GW data.
Even though we fix the phase transition parameters ($\delta_0$ and $\mu_C$ in this work), our analysis remains general enough. A Bayesian analysis within the CPS scenario without fixing the phase transition parameters results in smooth transitions only (see \ref{sec:var_PT_params}).

Additionally, we examine the parameter $d_c$, proposed in \cite{Annala:2023cwx} as a measure of conformality to assess the possibility of a quark core within NSs. Our findings suggest that NS cores may contain quark matter; however, within the current model, quark matter remains strongly interacting, and the conformal limit is not reached at the NS center. According to \cite{Annala:2023cwx}, models where $d_c$ falls below 0.2 indicate a transition to deconfined quark matter. In our calculations, we find that even with purely nucleonic degrees of freedom, $d_c$ gradually decreases below 0.2 around 5 $\rho_{B,0}$. For our hybrid model, it crosses this threshold at lower densities. This suggests that the threshold value for $d_c$, indicating deconfined quark matter, is not universally applicable across models.

Our results from Bayes factor calculation show a slight preference for hybrid EOSs when compared to NICER's mass-radius observations, particularly for pulsars PSR J0030+0451 and PSR J0740+6620. This suggests that quark matter within the EOS may better explain observed NS properties compared to purely nucleonic models. However, this preference is not conclusive, as GW170817 event data does not significantly distinguish between the two EOS models. The hybrid models, therefore, do not demonstrate clear superiority over nucleonic models based on current observational data.

We also examined the influence of NICER's latest mass-radius measurements of PSR J0437$-$4715 on the hybrid EOSs. We observe a decrease in the log-likelihood or Bayes factor (logarithmic scale), indicating a tension between earlier NICER data and the most recent measurements. This points to the need for more flexible EOS models, particularly hybrid ones, that can accommodate the constraints from both NICER and GW170817 observations. The observed inconsistency between NICER and LV datasets encourages studying more advanced EOS models to capture the diverse astrophysical phenomena associated with NSs.

\section{Acknowledgements}

AV would like to acknowledge the Human Resource Development Group, a division of CSIR (Council of Scientific \& Industrial Research, Ministry of Science \& Technology, Govt. of India) for the support through the CSIR-JRF 09/1026(16303)/2023-EMR-I.
TM acknowledges the support of RNCA (Rede Nacional de Computaçao Avançada), which is funded by the FCT (Fundaçao para a Ciência e a Tecnologia, IPP, Portugal). Furthermore, this work was completed with the assistance of Deucalion HPC, Portugal, under the Advanced Computing Project 2024.14108.CPCA.A3. T.M acknowledges support from FCT national funds for projects UIDB/04564/2020 (DOI: 10.54499/UIDB/04564/2020), UIDP/04564/2020 (DOI: 10.54499/UIDP/04564/2020), and 2022.06460.PTDC (DOI: 10.54499/2022.06460.PTDC).

\bibliographystyle{elsarticle-num} 
\bibliography{manuscript}

\begin{thebibliography}{10}
\expandafter\ifx\csname url\endcsname\relax
  \def\url#1{\texttt{#1}}\fi
\expandafter\ifx\csname urlprefix\endcsname\relax\def\urlprefix{URL }\fi
\expandafter\ifx\csname href\endcsname\relax
  \def\href#1#2{#2} \def\path#1{#1}\fi

\bibitem{Riley_2019}
T.~E. Riley, A.~L. Watts, S.~Bogdanov, et~al., Astrophys. J. Lett. 887 (2019) L21.

\bibitem{Miller_2019}
M.~C. Miller, F.~K. Lamb, A.~J. Dittmann, et~al., Astrophys. J. Lett. 887 (2019) L24.

\bibitem{Riley_2021}
T.~E. Riley, A.~L. Watts, P.~S. Ray, et~al., Astrophys. J. Lett. 918 (2021) L27.

\bibitem{Choudhury_2024}
D.~Choudhury, T.~Salmi, S.~Vinciguerra, et~al., arXiv:2407.06789 [astro-ph.HE] (2024).

\bibitem{Tews_2013}
I.~Tews, T.~Krüger, K.~Hebeler, et~al., Phys. Rev. Lett 110 (2013) 032504.

\bibitem{Drischler_2020}
C.~Drischler, R.~J. Furnstahl, J.~A. Melendez, et~al., Phys. Rev. Lett 125 (2020) 202702.

\bibitem{Huth_2021}
S.~Huth, C.~Wellenhofer, A.~Schwenk, Phys. Rev. C 103 (2021) 025803.

\bibitem{Malik_2023}
T.~Malik, M.~Ferreira, M.~B. Albino, et~al., Phys. Rev. D 107 (2023) 103018.

\bibitem{Venneti_2024}
A.~Venneti, S.~Gautam, S.~Banik, et~al., Phys. Lett. B 854 (2024) 138756.

\bibitem{LIGOScientific:2018cki}
B.~P. Abbott, R.~Abbott, T.~D. Abbott, et~al., Phys. Rev. Lett. 121 (2018) 161101.

\bibitem{Bauswein:2017vtn}
A.~Bauswein, O.~Just, H.-T. Janka, et~al., Astrophys. J. Lett. 850 (2017) L34.

\bibitem{Annala:2017llu}
E.~Annala, T.~Gorda, A.~Kurkela, et~al., Phys. Rev. Lett. 120 (2018) 172703.

\bibitem{Komoltsev2021jzg}
O.~Komoltsev, A.~Kurkela, Phys. Rev. Lett. 128 (2022) 202701.

\bibitem{Komoltsev_2024}
O.~Komoltsev, R.~Somasundaram, T.~Gorda, et~al., Phys. Rev. D 109 (2024) 094030.

\bibitem{Nandi_2019}
R.~Nandi, P.~Char, S.~Pal, Phys. Rev. C 99 (2019) 052802.

\bibitem{Biswas_2021}
B.~Biswas, P.~Char, R.~Nandi, et~al., Phys. Rev. D 103 (2021) 103015.

\bibitem{Ghosh_2022}
S.~Ghosh, D.~Chatterjee, J.~Schaffner-Bielich, The Eur Phys J A 58 (2022) 37.

\bibitem{Huth_2022}
S.~Huth, P.~T.~H. Pang, I.~Tews, et~al., Nature 606 (2022) 276.

\bibitem{Imam:2024gfh}
S.~M.~A. Imam, T.~Malik, C.~Provid\^encia, et~al., Phys. Rev. D 109 (2024) 103025.

\bibitem{Banik_2014}
S.~Banik, M.~Hempel, D.~Bandyopadhyay, Astrophys. J. Supp. Series 214 (2014) 22.

\bibitem{Chatterjee_2016}
D.~Chatterjee, I.~Vidaña, Eur. Phys. J A 52 (2016) 1.

\bibitem{Malik:2021nas}
T.~Malik, S.~Banik, D.~Bandyopadhyay, Astrophys. J. 910 (2021) 96.

\bibitem{Weber_2005}
F.~Weber, Prog. in Part. and Nuc. Phys. 54 (2005) 193.

\bibitem{Orsaria_2014}
M.~Orsaria, H.~Rodrigues, F.~Weber, et~al., Phys. Rev. C 89 (2014) 015806.

\bibitem{Malik_2021}
T.~Malik, S.~Banik, D.~Bandyopadhyay, Eur. Phys. J A 230 (2021) 561.

\bibitem{Alford:1997zt}
M.~G. Alford, K.~Rajagopal, F.~Wilczek, Phys. Lett. B 422 (1998) 247.

\bibitem{Bonanno:2011ch}
L.~Bonanno, A.~Sedrakian, Astron. Astrophys. 539 (2012) A16.

\bibitem{Ivanytskyi:2022oxv}
O.~Ivanytskyi, D.~Blaschke, Phys. Rev. D 105 (2022) 114042.

\bibitem{Mannarelli:2006fy}
M.~Mannarelli, K.~Rajagopal, R.~Sharma, Phys. Rev. D 73 (2006) 114012.

\bibitem{Rajagopal:2006ig}
K.~Rajagopal, R.~Sharma, Phys. Rev. D 74 (2006) 094019.

\bibitem{Haskell_2007}
B.~Haskell, N.~Andersson, D.~I. Jones, et~al., Phys. Rev. Lett 99 (2007) 231101.

\bibitem{Chodos:1974pn}
A.~Chodos, R.~L. Jaffe, K.~Johnson, C.~B. Thorn, Phys. Rev. D 10 (1974) 2599.

\bibitem{Johnson:1978uy}
K.~Johnson, Phys. Lett. B 78 (1978) 259.

\bibitem{Schertler:1999xn}
K.~Schertler, S.~Leupold, J.~Schaffner-Bielich, Phys. Rev. C 60 (1999) 025801.

\bibitem{Steiner:2000bi}
A.~Steiner, M.~Prakash, J.~M. Lattimer, Phys. Lett. B 486 (2000) 239.

\bibitem{Buballa:2003qv}
M.~Buballa, Phys. Rept. 407 (2005) 205.

\bibitem{Contrera:2022tqh}
G.~A. Contrera, D.~Blaschke, J.~P. Carlomagno, et~al., Phys. Rev. C 105 (2022) 045808.

\bibitem{Fogaca:2010mf}
D.~A. Fogaca, F.~S. Navarra, Phys. Lett. B 700 (2011) 236.

\bibitem{Tolos:2020aln}
L.~Tolos, L.~Fabbietti, Prog. Part. Nucl. Phys. 112 (2020) 103770.

\bibitem{Alford_2019}
M.~G. Alford, S.~Han, K.~Schwenzer, J. Phys. G: Nuc. and Part. Phys. 46 (2019) 114001.

\bibitem{Annala:2019puf}
E.~Annala, T.~Gorda, A.~Kurkela, et~al., Nat. Phys. 16 (2020) 907.

\bibitem{Annala:2023cwx}
E.~Annala, T.~Gorda, J.~Hirvonen, et~al., Nat. Commun. 14 (2023) 8451.

\bibitem{Andersson_1998}
N.~Andersson, K.~D. Kokkotas, Mon. Not. Roy. Astron. Soc. 299 (1998) 1059.

\bibitem{Kumar:2023rut}
D.~Kumar, T.~Malik, H.~Mishra, et~al., Phys. Rev. D 108 (2023) 083008.

\bibitem{Pradhan_2024}
B.~K. Pradhan, D.~Chatterjee, D.~E. Alvarez-Castillo, Mon. Not. Roy. Astron. Soc. 531 (2024) 4640.

\bibitem{Thakur_2024_arXiv}
P.~Thakur, S.~Chatterjee, K.~K. Nath, et~al., arXiv:2407.12601 [gr-qc] (2024).

\bibitem{Tan_2020}
H.~Tan, J.~Noronha-Hostler, N.~Yunes, Phys. Rev. Lett 125 (2020) 261104.

\bibitem{Shahrbaf_2020}
M.~Shahrbaf, D.~Blaschke, A.~G. Grunfeld, et~al., Phys. Rev. C 101 (2020) 025807.

\bibitem{Yamamoto_2023}
Y.~Yamamoto, N.~Yasutake, T.~A. Rijken, Phys. Rev. C 108 (2023) 035811.

\bibitem{Essick_2023}
R.~Essick, I.~Legred, K.~Chatziioannou, et~al., Phys. Rev. D 108 (2023) 043013.

\bibitem{Lin_2023}
Z.~Lin, A.~Steiner, arXiv:2310.01619 [astro-ph.HE] (2023).

\bibitem{GuhaRoy_2024}
D.~Guha~Roy, T.~Malik, S.~Bhattacharya, et~al., Astrophys. J 968 (2024) 124.

\bibitem{Typel_2015}
S.~Typel, M.~Oertel, T.~Klähn, Phys. of Part. and Nuc. 46 (2015) 633.

\bibitem{Albino:2024ymc}
M.~Albino, T.~Malik, M.~Ferreira, et~al., arXiv:2406.15337 [nucl-th] (2024).

\bibitem{Torres_2013}
J.~R. Torres, D.~P. Menezes, EPL 101 (2013) 42003.

\bibitem{Franzon:2012in}
B.~Franzon, D.~A. Fogaca, F.~S. Navarra, et~al., Phys. Rev. D 86 (2012) 065031.

\bibitem{Albino_2022}
M.~B. Albino, F.~S. Navarra, R.~Fariello, et~al., Jour. of Phys.: Conf. Series 2340 (2022) 012015.

\bibitem{Kapusta:2021ney}
J.~I. Kapusta, T.~Welle, Phys. Rev. C 104 (2021) L012801.

\bibitem{Constantinou:2023ged}
C.~Constantinou, T.~Zhao, S.~Han, M.~Prakash, Phys. Rev. D 107 (2023) 074013.

\bibitem{glendenning}
N.~K. Glendenning, Springer, 1997.

\bibitem{Hama_2006}
Y.~Hama, R.~P. Andrade, F.~Grassi, et~al., Nuc. Phys. A 774 (2006) 169.

\bibitem{Gorda_2023_link}
T.~Gorda, O.~Komoltsev, A.~Kurkela, {QCD} likelihood function, \url{https://zenodo.org/records/7781233} (2023).

\bibitem{Buchner_2014}
J.~Buchner, A.~Georgakakis, K.~Nandra, et~al., Astron. I\& Astrophys. 564 (2014) A125.

\bibitem{Lindblom_1983}
L.~Lindblom, S.~L. Detweiler, Astrophys. J. Supp. Series 53 (1983) 73.

\bibitem{Fackerell_1971}
E.~D. Fackerell, Astrophys. J. 166 (1971) 197.

\bibitem{Miller:2019cac}
M.~C. Miller, F.~K. Lamb, et~al., Astrophys. J. Lett. 887 (2019) L24.

\bibitem{Riley:2021pdl}
T.~E. Riley, et~al., Astrophys. J. Lett. 918 (2021) L27.

\bibitem{Miller:2021qha}
M.~C. Miller, et~al., Astrophys. J. Lett. 918 (2021) L28.

\bibitem{Fujimoto:2022ohj}
Y.~Fujimoto, K.~Fukushima, L.~D. McLerran, et~al., Phys. Rev. Lett. 129 (2022) 252702.

\bibitem{Miller_2021}
M.~C. Miller, F.~K. Lamb, A.~J. Dittmann, et~al., Astrophys. J. Lett. 918 (2021) L28.

\bibitem{LIGOScientific:2018hze}
B.~P. Abbott, et~al., Phys. Rev. X 9 (2019) 011001.

\bibitem{Typel_1999}
S.~Typel, H.~Wolter, Nuc. Phys. A 656 (1999) 331.

\bibitem{Dutra_2014}
M.~Dutra, O.~Lourenço, S.~S. Avancini, et~al., Phys. Rev. C 90 (2014) 055203.

\bibitem{Shlomo_2006}
S.~Shlomo, V.~M. Kolomietz, G.~Colò, Eur. Phys. J A 30 (2006) 23.

\bibitem{Todd_Rutel_2005}
B.~G. Todd-Rutel, J.~Piekarewicz, Phys. Rev. Lett 95 (2005) 122501.

\bibitem{Essick:2021ezp}
R.~Essick, P.~Landry, A.~Schwenk, et~al., Phys. Rev. C 104 (2021) 065804.

\bibitem{Hebeler:2013nza}
K.~Hebeler, J.~M. Lattimer, C.~J. Pethick, et~al., Astrophys. J. 773 (2013) 11.

\bibitem{jeffreys1998theory}
H.~Jeffreys, OuP Oxford, 1998.

\bibitem{Morey_2016}
R.~D. Morey, J.-W. Romeijn, J.~N. Rouder, J. Math. Psy. 72 (2016) 6.

\end{thebibliography}

\clearpage
\onecolumn
\section*{\Large SUPPLEMENTARY MATERIAL}

\appendix
We divide the supplementary material into six sections. The first one starts with a very brief intro to the Bayesian formalism. It is followed by the likelihood functions, astrophysical and nuclear physics constraints for the Bayesian inference, and citations for the data used.
The next section details Jeffrey's scale, which we follow to interpret the Bayes factor values for our EOS sets. We tabulate all the relevant data used to draw the conclusions for this work.
The third section contains results of couplings associated with the relativistic mean-field (RMF) model of the nucleonic equation of state (EOS), quark model parameters, nuclear matter parameters at nuclear saturation number density $\rho_{B,0}$, and neutron star (NS) observables like radius, maximum mass, $f-$mode oscillation frequencies and corresponding damping times. 
The fourth section justifies our choice of phase transition parameters $\delta_0$ and $\mu_C$ used in this work.
The fifth section lays out the theoretical basis for the relativistic mean-field model for hadronic EOSs and the mean-field model for the quark EOSs in this work.
The last section contains the results of statistical tests comparing \emph{Set0}, \emph{Set1} CPS and \emph{Set2} CPS.

\section{Likelihood functions, astrophysical and nuclear physics constraints for Bayesian inference}
\label{sec:loglikelihoods}

\begin{table*}[hb]
    \centering
    \setlength{\tabcolsep}{2.5pt}
\renewcommand{\arraystretch}{1.1}
    \begin{tabular}{cccccccc}
         \toprule
         Astrophysical Constraints & $M_\odot$ & R[km] & $\Tilde{\Lambda}$\footnotemark[8] & Refs. & \emph{Set0} & \emph{Set1} & \emph{Set2} \\
         \midrule
         PSR J0030+0451 (Riley) \footnotemark[1] & $1.34$ & $12.71^{+1.14}_{-1.19}$ & - & \cite{Riley_2019} & \checkmark & \checkmark & \checkmark \\
         PSR J0030+0451 (Miller) \footnotemark[2] & $1.44$ & $13.02^{+1.24}_{-1.06}$ & - & \cite{Miller_2019} & \checkmark & \checkmark & \checkmark \\
         \\
         PSR J0740+6620 (Riley) \footnotemark[3] & $2.07$ & $12.39^{+1.30}_{-0.98}$ & - & \cite{Riley_2021} & \checkmark & \checkmark & \checkmark \\
         PSR J0740+6620 (Miller) \footnotemark[4] & $2.08$ & $13.7^{+2.6}_{1.5}$ & - & \cite{Miller_2021} & \checkmark & \checkmark & \checkmark \\
         \\
         PSR J0437$-$4715 (Choudhury) \footnotemark[5] & $1.418$ & $11.36^{+0.95}_{-0.63}$ & - & \cite{Choudhury_2024} & $\times$ & $\times$ & \checkmark \\
         \\
         GW170817 (LIGO-Virgo) \footnotemark[6] & - & - & $300^{+420}_{-230}$ (HPD\footnotemark[7])& \cite{LIGOScientific:2018hze} & \checkmark & \checkmark & \checkmark \\
         \midrule
         \midrule
         Nuclear Physics Regime & Quantity & Constraints &  & Refs. & \emph{Set0} & \emph{Set1} & \emph{Set2} \\
         \midrule
         & $\rho_{B,0}$ & $0.153 \pm 0.005$ MeV &  & \cite{Typel_1999} &  \checkmark &  \checkmark &  \checkmark \\
         & $E_0$ & $-16.1 \pm 0.2$ MeV &  & \cite{Dutra_2014} & \checkmark & \checkmark & \checkmark \\
         SNM & $K_0$ & $230 \pm 40$ MeV &  & \cite{Shlomo_2006, Todd_Rutel_2005} & \checkmark & \checkmark & \checkmark \\
         & $J_0$ & $32.5 \pm 1.8$ MeV &  & \cite{Essick:2021ezp} & \checkmark & \checkmark & \checkmark \\
         \\
         & $P(\rho_B = 0.08 \text{ fm}^{-3})$ & $0.521 \pm 0.091$ MeV-fm$^{-3}$ &  &  & \checkmark & \checkmark & \checkmark \\
         PNM & $P(\rho_B = 0.12 \text{ fm}^{-3})$ & $1.262 \pm 0.295$ MeV-fm$^{-3}$ &  & \cite{Hebeler:2013nza} & \checkmark & \checkmark & \checkmark \\
         & $P(\rho_B = 0.16 \text{ fm}^{-3})$ & $2.513 \pm 0.675$ MeV-fm$^{-3}$
         &  &  & \checkmark & \checkmark & \checkmark \\
         \\
         pQCD & EOS at $\rho_B = 7 \times \rho_{B,0}$ & as given in \cite{Gorda_2023_link} &  &\cite{Komoltsev2021jzg} & \checkmark & \checkmark & \checkmark \\
         \bottomrule
    \end{tabular}
    \caption{Astrophysical and nuclear physics constraints used during Bayesian inference of the model parameters to generate nucleonic and hybrid EOS sets employed in this work. \emph{Set1} and \emph{Set2} in the columns refer to the corresponding sets within Fixed Hadronic and Combined Parameter Sampling scenarios. SNM, PNM, and pQCD refer to Symmetric Nuclear Matter, Pure Neutron Matter, and perturbative Quantum Chromodynamics. $P$ is pressure. $\rho_{B,0}$, $E_0$, $K_0$, and $J_0$ stand for baryon number density, energy per baryon, incompressibility coefficient, and symmetry energy, respectively, calculated at nuclear saturation. }
    \label{tab:astro_nuc_data}
\end{table*}

\footnotetext[1]{https://zenodo.org/records/8239000}
\footnotetext[2]{https://zenodo.org/records/3473466}
\footnotetext[3]{https://zenodo.org/records/4697625}
\footnotetext[4]{https://zenodo.org/records/4670689}
\footnotetext[5]{https://zenodo.org/records/12703175}
\footnotetext[6]{https://dcc.ligo.org/LIGO-P1800115/public}
\footnotetext[7]{Highest posterior density}
\footnotetext[8]{Reduced tidal deformability parameter}

The Bayesian approach for parameter estimation allows us to constrain the parameter space associated with a model by integrating prior beliefs about the parameters with new data. Incorporating different datasets during Bayesian inference can reveal interesting correlations among parameters in the posterior distribution which may not be obvious otherwise.
The posterior distribution for the model parameters $\theta$ in Bayes' theorem is given as
\begin{equation}
    P(\mathbf{\theta} \vert D) = \dfrac{\mathcal{L}(D \vert \mathbf{\theta}) P(\theta)}{\mathcal{Z}}
\end{equation}
with $\theta$ and $D$ denoting the set of model parameters and the fit data, respectively.

\subsection{Chiral effective field theory constraints}
\begin{equation}
    \mathcal{L}^{PNM}(D \vert \theta) = \prod_j \frac{1}{2 \sigma^2_j} \dfrac{1}{\exp \left( \dfrac{\vert d_j - m_j(\theta) \vert - \sigma_j}{0.015} \right) + 1}
\end{equation}
where $d_j$ is the median value and $\sigma_j$ represents two times the uncertainty of the $j^{th}$ data point from the $\chi$EFT constraints \cite{Hebeler:2013nza}.

\subsection{pQCD constraints}
\begin{equation}
    \mathcal{L} (d_{pQCD} \vert \theta) = P(d_{pQCD} \vert \theta) = \mathcal{L}^{pQCD},
\end{equation}
where $d_{pQCD}$ is the probability distribution in the energy density and pressure plane at 7 times the nuclear saturation density. Please refer to the references listed in Table \ref{tab:astro_nuc_data} for details on the implementation. We enforce the pQCD constraints stringently, meaning that we can require the squared speed of sound to be below 1/3 from the given point up to the specified scale $\mu_{\rm QCD}$, setting $\mu_{\rm QCD}$ to 2.4 GeV rather than the usual 2.6 GeV, as mentioned in \cite{Komoltsev2021jzg}. 
\subsection{Gravitational wave and x-ray observation constraints}
\begin{align}
    P(d_{GW} \vert EOS) = &\int^{M_u}_{m_2} dm_1 \int^{m_1}_{M_l} dm_2 P(m_1, m_2 \vert EOS) \nonumber\\
    &\times P(d_{GW} \vert m_1, m_2, \Lambda_1(m_1, EOS)) \nonumber\\
    &\Lambda_2 (m_2, EOS)) \nonumber\\
    & = \mathcal{L}^{GW},
\end{align}
where $P(m \vert EOS)$ is given by

\begin{equation}
    P(m \vert EOS) =
    \begin{cases}
         &\dfrac{1}{M_u - M_l}\text{    if }M_l \leq m \leq M_u,\\
         &0\text{      else}.
    \end{cases}
\end{equation}

We can determine the mass and radius from the data for X-ray observation by NICER. The corresponding evidence takes the form:
 \begin{align}
     P(d_{x-ray} \vert EOS) = &\int^{M_u}_{M_l} dm P(m \vert EOS) \nonumber\\
     &\times P(d_{x-ray} \vert m, R(m, EOS)) \nonumber\\
     &= \mathcal{L}^{NICER}.
 \end{align}
where $M_l = \text{1} M_\odot$, and $M_u$ denotes the maximum mass of a NS for a given EOS.

\subsection{Phase transition likelihood function}
\label{sec:PT_llhood}
We use the following supergaussian to calculate the likelihood of the transition number density inside the window for $\rho_{B,tr}$:\\
\begin{equation}
    \mathcal{L}^{PT} = \exp \left[ (\rho_{B,tr} - \mu)^2 / (2\sigma^2)^3 \right]
\end{equation}
where $\sigma=0.07$ and $\mu=2.1 \times \rho_{B,0}$ with $\rho_{B,0}=0.16$ fm$^{-3}$.

\subsection{Total likelihood function}
We can write the combined likelihood function for the different sets in the CPS scenario to be:
\begin{equation}
    \mathcal{L}^{total}_{Set0} = \mathcal{L}^{PNM} \times \mathcal{L}^{GW} \times \mathcal{L}^{NICER~I} \times  \mathcal{L}^{NICER~II} \times \mathcal{L}^{pQCD}
\end{equation}

\begin{equation}
    \mathcal{L}^{total}_{Set1} = \mathcal{L}^{PNM} \times \mathcal{L}^{GW} \times \mathcal{L}^{NICER~I} \times  \mathcal{L}^{NICER~II} \times \mathcal{L}^{PT} \times \mathcal{L}^{pQCD}
\end{equation}

\begin{equation}
    \mathcal{L}^{total}_{Set2} = \mathcal{L}^{PNM} \times \mathcal{L}^{GW} \times \mathcal{L}^{NICER~I} \times  \mathcal{L}^{NICER~II} \times  \mathcal{L}^{NICER~III} \times \mathcal{L}^{PT} \times \mathcal{L}^{pQCD}
\end{equation}

$NICER~I$, $NICER~II$, and $NICER~III$ correspond to the mass-radius measurements of PSR J0031+0451, PSR J0740+6620, and PSR J0437$-$4715 respectively.

\section{Bayes factor calculation}
\label{tab:Jeffreys}
Jeffrey's scale\cite{jeffreys1998theory, Morey_2016}
used to interpret the strength of the evidence provided by Bayes factors is given as:
\begin{itemize}[noitemsep,nolistsep]
    \item
    \textbf{0} to \textbf{1}: Minimal evidence
    \item
    \textbf{1} to \textbf{2.5}: Substantial evidence
    \item
    \textbf{2.5} to \textbf{5}: Strong evidence
    \item
    \textbf{$>$5} : Decisive evidence
\end{itemize}

\begin{figure*}
    \centering
    \includegraphics[width=1\linewidth]{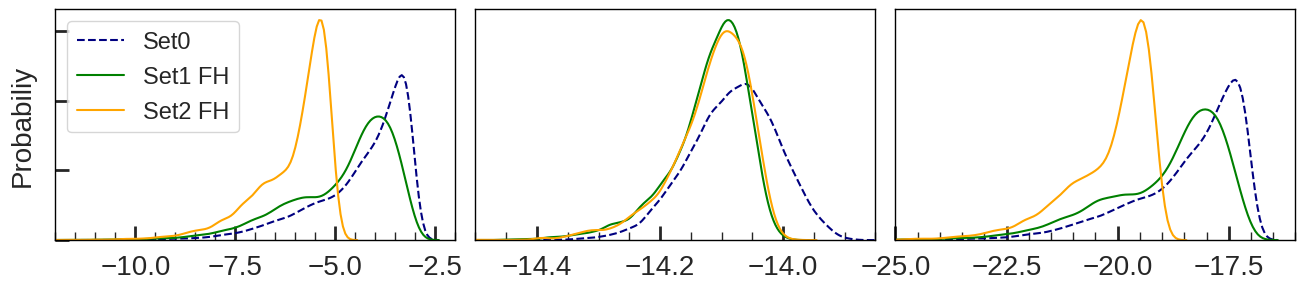}
    \includegraphics[width=1\linewidth]{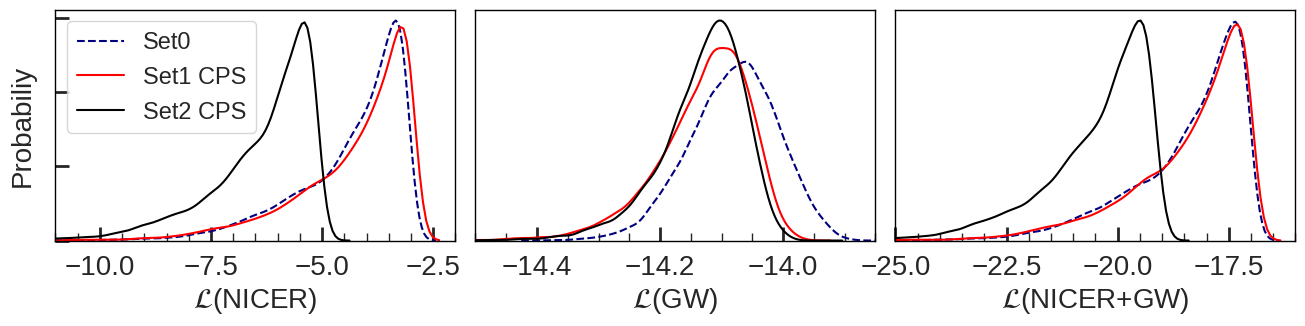}
    \caption{Probability distribution of log-likelihoods associated with NICER, GW, and NICER+GW constraints for the posterior distributions of EOSs. The top panel presents the log-likelihoods for \emph{Set0}, \emph{Set1} FH, and \emph{Set2} FH, and the bottom panel illustrates the same for \emph{Set0}, \emph{Set1} CPS and \emph{Set2} CPS. The position of the peaks differs when we calculate log-likelihood for NICER observations only.}
    \label{fig:log_likelihoods}
\end{figure*}

\begin{table}[ht]
\centering
\begin{tabular}{lccc}
    \toprule
    \textbf{Set} & \textbf{NICER} & \textbf{GW} & \textbf{NICER+GW} \\
    \midrule
    \emph{Set0}     & -3.89 & -14.08 & -17.93 \\
    & & & \\
    \emph{Set1} FH  & -4.25 & -14.12 & -18.35 \\
    \emph{Set2} FH  & -5.76 & -14.12 & -19.85 \\
    & & & \\
    \emph{Set1} CPS & -3.68 & -14.13 & -17.78 \\
    \emph{Set2} CPS & -5.90 & -14.13 & -20.00 \\
    \bottomrule
    \end{tabular}
    \caption{Estimated Bayes Log Evidence for different sets and constraints}
    \label{tab:bayes_log_evidence}
\end{table}

\begin{table}
\centering
\caption{Bayes Factors (Log Scale) among different sets for NICER, GW, and NICER+GW}
\begin{tabular}{lccc}
\toprule
\textbf{Comparison} & \textbf{NICER} & \textbf{GW} & \textbf{NICER+GW} \\
\midrule
\emph{Set0} vs. \emph{Set1} FH & $-0.36$ & $-0.04$ & $-0.42$ \\
\emph{Set0} vs. \emph{Set2} FH & $-1.87$ & $-0.04$ & $-1.92$ \\
\emph{Set0} vs. \emph{Set1} CPS & $+0.21$ & $-0.05$ & $+0.15$ \\
\emph{Set0} vs. \emph{Set2} CPS & $-2.01$ & $-0.05$ & $-2.07$ \\
\emph{Set1} FH vs. \emph{Set2} FH & $-1.51$ & $0.00$ & $-1.50$ \\
\emph{Set1} FH vs. \emph{Set1} CPS & $+0.57$ & $-0.01$ & $+0.57$ \\
\emph{Set1} FH vs. \emph{Set2} CPS & $-1.65$ & $-0.01$ & $-1.65$ \\
\emph{Set2} FH vs. \emph{Set1} CPS & $+2.08$ & $-0.01$ & $+2.07$ \\
\emph{Set2} FH vs. \emph{Set2} CPS & $-0.14$ & $-0.01$ & $-0.15$ \\
\emph{Set1} CPS vs. \emph{Set2} CPS & $-2.22$ & $0.00$ & $-2.22$ \\
\bottomrule
\end{tabular}
\label{tab:bayes_factors}
\end{table}

\newpage
\section{Posterior distribution of nuclear matter parameters and neutron star observables}
\begin{figure*}
    \centering
    \includegraphics[width=0.9\linewidth]{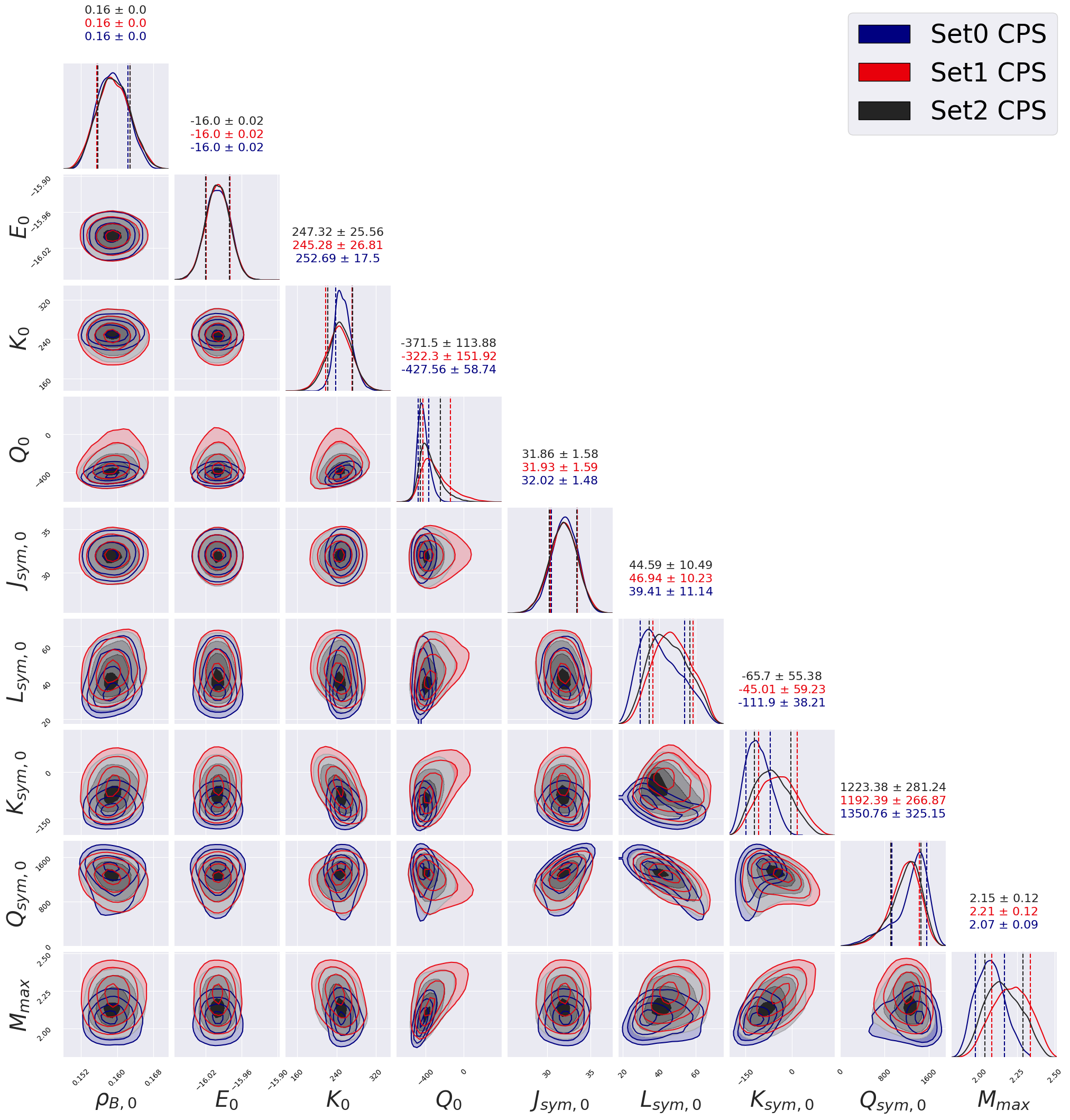}
    \caption{Corner plot of posterior distributions of some nuclear matter parameters (NMPs) like saturation number density ($\rho_{B,0}$), the binding energy per nucleon ($E_0$), incompressibility coefficient ($K_0$), the skewness coefficient ($Q_0$), symmetry energy ($J_{sym,0}$), the curvature of symmetry energy ($K_{sym,0}$), the slope of symmetry energy ($L_{sym,0}$), and the skewness of symmetry energy ($Q_{sym,0}$), calculated at the saturation number density. Except for $\rho_{B,0}$, which is in fm$^{-3}$, the values of all the other NMPs are in MeV. The posterior distribution of maximum mass ($M_{max}$) in $M_\odot$ of a given EOS is also plotted. Vertical dashed lines over the panels with the marginalized 1D distribution of each quantity delineate a given quantity's 1$\sigma$ credible interval (CI). The median values for all the quantities are reported with the bounds for 1$\sigma$ (68\%) CI.}
    \label{fig:corner_NMP}
\end{figure*}

\begin{table*}
    \centering
    \begin{tabular}{ccccccc}
        \toprule
        & \multicolumn{2}{c}{\emph{Set0}} & \multicolumn{2}{c}{\emph{Set1} CPS} & \multicolumn{2}{c}{\emph{Set2} CPS} \\
        \midrule
        Quantity & Med. & 90\% CI & Med. & 90\% CI & Med. & 90\% CI \\
        \midrule
        $g_\sigma$ & 8.406 & [7.852, 8.924] & 9.171 & [8.210, 9.920] & 8.944 & [8.090, 9.717]\\
        $g_\omega$ & 9.907 & [8.830, 10.834] & 11.337 & [9.619, 12.613] & 10.927 & [9.363, 12.257] \\
        $g_\rho$ & 11.530 & [9.117, 13.941] & 11.574 & [8.972, 14.112] & 11.544 & [8.998, 14.080] \\
        $B$ & 4.730 & [3.315, 7.097] & 3.114 & [2.160, 4.922] & 3.493 & [2.310, 5.442] \\
        $C$ & -2.965 & [-4.722, 2.774] & -2.882 & [-4.475, 0.026] & -2.907 & [-4.600, 0.885] \\
        \\
        $\xi$ & 0.0038 & [0.0004, 0.0111] & 0.006 & [0.001, 0.014] & 0.0057 & [0.0007, 0.0154] \\
        $\Lambda_\omega$ & 0.064 & [0.037, 0.105] & 0.046 & [0.030, 0.076] & 0.050 & [0.031, 0.086] \\
        \\
        $\rho_{B,0}$ & 0.159 & [0.154, 0.164] & 0.159 & [0.153, 0.165] & 0.159 & [0.153, 0.165] \\
        $E_0$ & -16.00 & [-16.03, -15.97] & -16.00 & [-16.03, -15.97] & -16.00 & [-16.03, -15.97] \\
        $K_0$ & 252.69 & [228.79, 285.35] & 245.28 & [201.02, 288.86] & 247.32 & [204.20, 289.39] \\
        $Q_0$ & -427.56 & [-500.93, -312.24] & -322.30 & [-483.33, 11.98] & -371.50 & [-495.82, -128.71] \\
        $Z_0$ & 2088 & [666, 3509] & 4014 & [871, 9489] & 3166 & [735, 7465] \\
        \\
        $J_{sym,0}$ & 32.02 & [29.51, 34.33] & 31.93 & [29.22, 34.49] & 31.86 & [29.19, 34.40] \\
        $L_{sym,0}$ & 39.41 & [25.41, 61.54] & 46.94 & [30.89, 64.54] & 44.59 & [29.22, 63.41] \\
        $K_{sym,0}$ & -111.90 & [-167.74, -42.68] & -45.01 & [-140.64, 55.01] & -65.70 & [-148.81, 33.78] \\
        $Q_{sym,0}$ & 1351 & [585, 1661] & 1192 & [684, 1555] & 1223 & [650, 1581] \\
        $Z_{sym,0}$ & -11081 & [-18601, 231] & -13458 & [-18888, -1539] & -12752 & [-18583, -983] \\
        \\
        $\xi_{quark}$ & - & - & 0.0013 & [0.0011, 0.0014] & 0.0012 & [0.0010, 0.0014] \\
        $B_{quark}$ & - & - & 74.45 & [69.35, 80.37] & 75.27 & [70.50, 80.68]\\
        \\
        $M_{max}$ & 2.067 & [1.924, 2.232] & 2.213 & [2.009, 2.406] & 2.151 & [1.972, 2.356] \\
        $R_{1.4}$ & 12.26 & [11.85, 12.63] & 2.51 & [11.92, 13.07] & 12.33 & [11.79, 12.84] \\
        $\Lambda_{1.4}$ & 389 & [305, 480] & 447 & [319, 614] & 403 & [295, 540] \\
        \\
        $\nu_{f_{1.4}}$ & 1.812 & [1.731, 1.907] & 1.712 & [1.616, 1.824] & 1.744 & [1.649, 1.851] \\
        $\tau_{f_{1.4}}$ & 0.111 & [0.101, 0.122] & 0.124 & [0.110, 0.140] & 0.120 & [0.107, 0.134] \\
        \bottomrule
    \end{tabular}
    \caption{Table of median values and 90\% credible interval values from the posterior distribution of RMF model parameters($g_\sigma$, $g_\omega$, $g_\rho$, $B$, $C$, $\xi$, $\Lambda_\omega$), MFTQCD model parameters ($\xi_{quark}$ and $B_{quark}$), nuclear saturation properties ($\rho_{B,0}$, $E_0$, $K_0$, $Q_0$, $Z_0$, $J_{sym,0}$, $L_{sym,0}$, $K_{sym,0}$, $Q_{sym,0}$, $Z_{sym,0}$), and NS observables ($M_{max}$, $R_{1.4}$, $\Lambda_{1.4}$, $\nu_{f_{1.4}}$, $\tau_{f_{1.4}}$) for the combined parameter sampling (CPS) scenario. $B$ and $C$ are defined as $b \times 10^3$ and $c \times 10^3$ - $b$ and $c$ represent the strength of non-linear terms in the RMF Lagrangian\cite{Malik_2023}. All the RMF model parameters are dimensionless. $\xi_{quark}$ and $B_{quark}$ are in MeV$^{-1}$ and MeV/fm$^{-3}$. All the nuclear matter parameters are in MeV except $\rho_{B,0}$, which is in fm$^{-3}$. $M_{max}$, $R_{1.4}$, $\nu_{f_{1.4}}$, and $\tau_{f_{1.4}}$ are in $M_\odot$, km, kHz, and seconds while $\Lambda_{1.4}$ is dimensionless.}
    \label{tab:post_CPS}
\end{table*}

\clearpage
\section{Choice of phase transition parameters}
We intend to avoid any preference in favour of Gibbs or Maxwell transitions. Hence, we choose to construct the phase transitions in our hybrid EOSs using $(P-P_Q)(P-P_H)=\delta(\mu_B)= \delta_0 \exp [-(\mu_B/\mu_c)]$, where the $P_H$ and $P_Q$ are pressure for hadronic and quark phases respectively \cite{Albino_2022, Hama_2006}. The smoothness of the phase transitions are dictated by two free parameters, $\delta_0$ and $\mu_C$. $\delta_0=0$ (MeV/fm$^3$)$^2$ and/or $\mu_C \approx 0 - 200$ MeV corresponds to EOSs with a sharp discontinuity in energy density; higher values of $\delta_0$ result in EOSs with smoother transitions \cite{Albino_2022}.

\begin{figure}[h]
    \centering
    \includegraphics[width=0.8\linewidth]{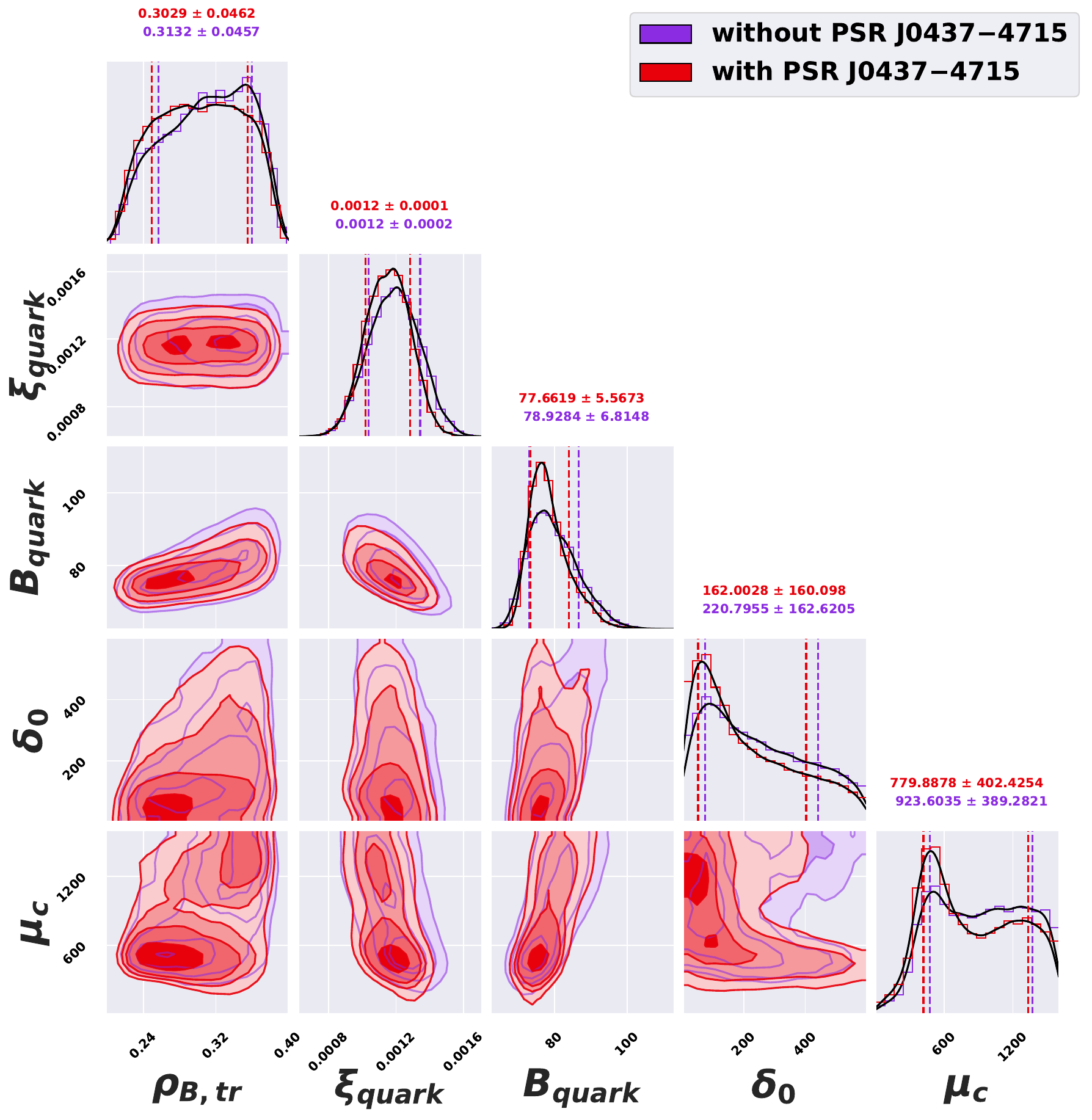}
    \caption{Corner plot of transition number density ($\rho_{B,tr}$), MFTQCD quark EOS parameters --- $\xi_{quark}$ and $B_{quark}$, and phase transition parameters --- $\delta_0$, and $\mu_C$ for a Bayesian sampling within the CPS scenario involving variation of $\delta_0$, and $\mu_C$. Red and blue contours, dashed lines and median values represent data from posteriors obtained with and without PSR J0437$-$4715 data among the NICER constraints, respectively. The vertical dashed lines mark the 1$\sigma$ (68\%) credible interval bounds for a given quantity.}
    \label{fig:corner_PT_params}
\end{figure}

In order to keep our analysis as general as possible, we must define prior ranges for the parameters controlling the smoothness of phase transitions. Fig. \ref{fig:corner_PT_params} containing the results of an analysis within the CPS scenario without fixing $\delta_0$ and $\mu_C$, suggest that our likelihood evaluations do not constrain them well enough and have a slight preference for $\delta_0 = 50$ (MeV/fm$^3$)$^2$ and $\mu_C = 700$ MeV. But we find one full-fledged execution of such analysis to be computationally exorbitant. Thus, we choose the aforementioned values to perform a less resource-intensive and quicker Bayesian analysis without any loss of generality.
\label{sec:var_PT_params}

\newpage
\section{Formalism for hadronic and quark equations of state}
\label{sec:hadron_quark_EOS}

The Lagrangian for the relativistic mean-field (RMF) model\cite{GuhaRoy_2024} includes multiple non-linear meson interaction terms to represent the hadronic EOS. The nuclear interaction between nucleons is introduced through the exchange of the scalar-isoscalar meson $\sigma$, the vector-isoscalar meson $\omega$, and the vector-isovector meson $\varrho$. 

The Lagrangian density is written as \cite{Dutra_2014, Malik_2023}
\begin{equation}
  \mathcal{L}=   \mathcal{L}_N+ \mathcal{L}_M + \mathcal{L}_{NL} +\mathcal{L}_{leptons},
\end{equation} 
where
\begin{equation}
\mathcal{L}_{N} = \bar{\Psi}\Big[\gamma^{\mu}\left(i \partial_{\mu}-g_{\omega} \omega_\mu - \frac{1}{2}g_{\varrho} {\boldsymbol{t}} \cdot \boldsymbol{\varrho}_{\mu}\right) - \left(m_N - g_{\sigma} \sigma\right)\Big] \Psi,
\end{equation}
represents the Dirac equation for the nucleon doublet (neutron and proton) with bare mass $m_N$, $\Psi$ is a Dirac spinor, $\gamma^\mu $ are the Dirac matrices, and $\boldsymbol{t}$ is the isospin operator. $\mathcal{L}_{M}$ denotes the mesons, given by
\begin{equation}
\mathcal{L}_{M}  = \frac{1}{2}\left[\partial_{\mu} \sigma \partial^{\mu} \sigma-m_{\sigma}^{2} \sigma^{2} \right] - \frac{1}{4} F_{\mu \nu}^{(\omega)} F^{(\omega) \mu \nu} + \frac{1}{2}m_{\omega}^{2} \omega_{\mu} \omega^{\mu} - \frac{1}{4} \boldsymbol{F}_{\mu \nu}^{(\varrho)} \cdot \boldsymbol{F}^{(\varrho) \mu \nu} + \frac{1}{2} m_{\varrho}^{2} \boldsymbol{\varrho}_{\mu} \cdot \boldsymbol{\varrho}^{\mu}
\end{equation}
where $F^{(\omega, \varrho)\mu \nu} = \partial^ \mu A^{(\omega, \varrho)\nu} -\partial^ \nu A^{(\omega, \varrho) \mu}$ are the vector meson  tensors.
\begin{equation}
\mathcal{L}_{NL} = -\frac{1}{3} b~m_N~ g_\sigma^3 (\sigma)^{3}-\frac{1}{4} c (g_\sigma \sigma)^{4}+\frac{\xi}{4!} g_{\omega}^4 (\omega_{\mu}\omega^{\mu})^{2} + \Lambda_{\omega}g_{\varrho}^{2}\boldsymbol{\varrho}_{\mu} \cdot \boldsymbol{\varrho}^{\mu} g_{\omega}^{2}\omega_{\mu}\omega^{\mu}
\end{equation}
involves the non-linear terms with parameters $b$, $c$, $\xi$, $\Lambda_{\omega}$ to take care of the high-density behaviour of the matter. 
$g_i$'s are the couplings of the nucleons to the meson fields $i = \sigma, \omega, \varrho$, with masses $m_i$.
\noindent
Finally, the Lagrangian density for the non-interacting leptons is written as
\begin{equation}
\mathcal{L}_{leptons}= \bar{\Psi_l}\Big[\gamma^{\mu}\left(i \partial_{\mu}  
-m_l \right)\Psi_l\Big],
\end{equation}
where $\Psi_l~(l= e^-, \mu^-)$ denotes the lepton spinor;

\vspace{0.5cm}

The MFTQCD formalism provides us with the quark EOSs \cite{Albino:2024ymc} from the effective Lagrangian\cite{Fogaca:2010mf}
\begin{equation}
\label{eq:MFTQCD}
\mathcal{L} = -b\phi_0^4 + \frac{m^2_G}{2} \alpha^a_0 \alpha^a_0 + \bar{\psi}^q_i (i \delta_{ij} \gamma^\mu \partial_\mu + g_h \gamma^0 T^a_{ij} \alpha^a_0 - \delta_{ij}m)\psi^q_j
\end{equation}
where $\xi_{quark} = g/m_G$, $g$ is strong coupling constant, $m_G$ is dynamical gluon mass, and $b\phi_0^4$ is similar to the bag constant of MIT Bag model.
$T^a$ are the SU(3) generators and $g_h$ is the interaction between the high momentum component $\alpha^a_0$ of the gluon field and the quark of ﬂavor $q$. The $i$ and $j$ are the colour indices of the quarks.

In this formalism, the gluon field is assumed to consist of two components, soft and hard, corresponding to low and high momentum, respectively.
The Lagrangian in Eq. \ref{eq:MFTQCD} characterizes a system with quarks, soft gluons and hard gluons\cite{Fogaca:2010mf}. The quarks couple only to the hard gluons. The hard gluons couple to the quarks and to the soft gluons. The hard gluon ﬁeld ($\alpha^{a \mu}$) propagates without exchanging momenta on a soft gluon ﬁeld ($A^{a \mu}$) background. The hard gluon gains an effective mass as a result of this interaction.

\section{Statistical tests comparing \emph{Set0}, \emph{Set1} CPS and \emph{Set2} CPS}
\label{sec:stat_tests}
Kullback-Leibler (KL) divergence, is simply a measure of the difference between two given PDFs. 
The KL divergence values are 2.130, 1.729, and 1.909 for \emph{Set0}, \emph{Set1} CPS, and \emph{Set2} CPS, respectively.\\
We use the Kolmogorov-Smirnov (KS) test to compare two given cumulative distribution functions (CDFs) corresponding to the PDFs. The KS test statistic values with respect to the D$_L$ for the NICER data are 0.912, 0.852, and 0.878, respectively, for \emph{Set0}, \emph{Set1} CPS, and \emph{Set2} CPS, respectively.

\end{document}